\documentclass[aps,prl,twocolumn]{revtex4}
\usepackage{amsmath}
\usepackage{graphicx}
\usepackage{dsfont}

\begin{document}

\title{Plaquette-centered rotation symmetry and octet-nodal
superconductivity in $\text{KFe}_{2}\text{As}_{2}$}
\author{Guo-Yi Zhu$^{1}$ and Guang-Ming Zhang$^{1,2}$}
\affiliation{$^{1}$State Key Laboratory of Low-Dimensional Quantum Physics and Department
of Physics, Tsinghua University, Beijing 100084, China. \\
$^{2}$Collaborative Innovation Center of Quantum Matter, Beijing 100084,
China.}
\date{\today}

\begin{abstract}
A plaquette-centered rotation symmetry $C_{4}^{p}$ is identified to play a
significant role in determining and stabilizing the Fermi-surface structure
of Fe-based superconductors. Together with the $S_{4}$ symmetry previously
found, we are able to sort out the tangling orbitals and solve the puzzle of
pairing symmetry of superconductivity in $\text{KFe}_{2}\text{As}_{2}$ in a
simple but comprehensive way. By modeling the material with a strong
coupling $t-J_{1}-J_{2}$ model, we find phase transitions of pairing
symmetry driven by the competition between the local spin antiferromagnetic
couplings from nodal $d_{x^2-y^2}\times s_{x^2+y^2}$-wave to nodeless $%
s_{x^2 y^2} $-wave through the intermediate $s+id\times s$ mixed pairing
phase, which is consistent with the observation of pressure experiments. The
emergent $d$-wave form factor inevitably arises from the projection of
inter-orbital Cooper pairing onto the Fermi surface and is inherited from
the electronic structure in the representation of $C_{4}^{p}$ symmetry.
Moreover, the $S_{4}$ symmetry dictates 2 copies of $d$-wave pairing
condensates, counting 8 nodes in total. We further show that weakly breaking
$C_{4}^{p}$ naturally leads to the octet nodal gap as precisely observed in
laser angle resolved photoemission spectroscopy. The octet nodes reflect the
collaboration of the $C_{4}^{p}$ and $S_{4}$ symmetries, which sheds new
light on the enigma of the pairing symmetry in $\text{KFe}_{2}\text{As}_{2}$.
\end{abstract}

\maketitle

As the first family of high temperature superconductivity (SC), the cuprates
have been under elaborate studies for nearly thirty years\cite%
{Muller,Anderson,AtoZ,LeeNagaosaWen}, and a consensus has been reached by
the majority of condensed matter physicists that the intriguing
superconducting phase of cuprates are rooted in the electronic strong
antiferromagnetic (AF) correlation. The SC in optimally doped compounds is
confirmed by a vast majority experiments to be $d$-wave pairing symmetry\cite%
{Shen,Harlingen,Tsuei}. For the theorists, the complex reality arising from
both the $3d$ orbital on Cu ions and $2p$ orbital on O ions is compromised
by formation of the Zhang-Rice singlets, which justifies a rather successful
description of a single-band model for the cuprates - the celebrated t-J
model\cite{ZhangRice,KotliarLiu,Joynt,AtoZ,LeeNagaosaWen}.

The phase diagrams of Fe-based superconductors show significant similarity
with those of the cuprates, suggesting deep connection of the essential
physics between them. However, distinct from the cuprates, the family of
Fe-based superconductors has a seemingly less unified picture as to their
Fermi surface (FS) structure, pairing symmetries and pairing glues\cite%
{Review1,Review2,Review3,Review4,Review5,Review6}. From the perspective of
band structure, they owe their complexity to the multi-orbital character.
Some Fermi pockets could vanish under doping or pressure and are less robust
than the others\cite{SatoDingPrl2009,UjiFSmassEnhance}. Regarding the
pairing symmetry, the $s_{\pm }$-wave pairing symmetry has been established
for many iron-pnictides from different approaches\cite%
{HuBernevig2orbitalExchange,scalapino2009,YuZhuSi2014}, but it fails to
account for all the materials observed experimentally. When it comes to the
interactions responsible for pairing, there are weak coupling theories that
count on the antiferromagnetic spin fluctuation\cite{scalapino2009}, but
they are less satisfying when FS nesting is absent, nor when there exists
strong correlation between local magnetic moments\cite%
{Review1,Review2,Review3,Review4,Review5,Review6}. After all these
considerations, a simple but unifying theoretical recipe for the Fe-based
superconductors is still elusive, and we still need deep reflection upon the
electronic structure and interactions of Fe-based superconductors.

To model the Fe-based SCs, we have to firstly understand the electronic
structure. There have been many theoretical treatments by taking all five or
even ten $3d$-orbitals of Fe into consideration, but the necessity for such
a huge parameter space is hardly conceivable, because the pairing symmetries
are so robust. From this point of view, we especially appreciate the $S_{4}$
symmetry found by Hu and Hao\cite{S4HuHao2012}, which provides a rational
way to sort out the tangling multi-orbitals of Fe-based SCs and reduce the
parameters of the electronic structure. In the presence of $S_{4}$, the band
structure of multi-orbitals is then reduced to a minimal two-band model.
Based on this $S_{4}$ symmetry, the model with a few parameters is capable
of accommodating most FS structures of both iron-pnictides and
iron-chalcogenides\cite{S4HuHao2012}, potentially unifying the electronic
structure of Fe-based SCs. However, the heavily hole doped limit of
iron-pnictides $\text{KFe}_{2}\text{As}_{2}$ is left behind.

The representative of heavily hole doped Ba$_{1-x}$K$_{x}$Fe$_{2}$As$_{2}$
family - $\text{KFe}_{2}\text{As}_{2}$ is a particular intriguing case among
the Fe-based superconductors, regarding its pairing symmetry and
interaction. Angle resolved photoemission spectroscopy (ARPES) experiments%
\cite{SatoDingPrl2009,UjiFSmassEnhance} report the absence of electron
pockets and leave only three hole Fermi pockets around the Brillouin zone
(BZ) center ($\Gamma $ point). Usually the nodeless $s$-wave pairing
symmetry is favorable. But it turns out that nodal gap structure was
observed by thermal conductivity measurement\cite{SYLiPrl2010}, penetration
depth measurement\cite{HashimotoPrb2010}, and NMR probe\cite%
{Fukazawa2009,WYuPrb2010,wutao2016}. The $d$-wave pairing symmetry was then
proposed after functional renormalization group calculation\cite%
{sdWaveHuBernevig2011}, which seems consistent with the measurement of heat
conduction\cite{dWaveTaillefer2012}. Later on Okazaki et. al. utilized the
ultra-high energy resolution of laser ARPES to map the gap structure much
more explicitly\cite{OctetOkazaki2012}. Quite dramatically, their report
showed octet nodal gap on the middle hole pocket while nodeless gap on the
inner hole pocket, in striking contrast to the theoretical expectation. Then
there were attempts to resort to accidental nodes to account for the octet
nodes. But as far as we are concerned, the nodes robustly observed in
various experiments deserve far more convincing explanation. If the octet
nodes was given by an effective g-wave pairing condensate
phenomenologically, it is a natural question how the Cooper pair obtains
angular momentum as high as $l=4$ microscopically. Although the so-called
concealed d-wave scenario was proposed to explain the formation of the
effective g-wave Cooper pair, such a theory starts from an orbital selective
pairing form, and their orbital triplet pairing vector field is required to
be locked with the orbital Rashba vector field and to rotate oppositely\cite%
{Coleman}.

In this paper, we will provide a mechanism based on a microscopic minimal
model, which is much simpler but more transparent. Our idea is inspired by
another experiment. Taillefer's group reported their pressure study on $%
\text{KFe}_{2}\text{As}_{2}$, in which they witnessed a non-monotonic
V-shaped tendency of the superconducting transition temperature with
increasing pressure\cite{TaftiTaillefer2013}. Their results clearly
indicated a phase transition near the point of the sudden change of $T_{c}$.
The possibility of FS transition was ruled out by their Hall coefficient
measurement, pointing towards a phase transition of pairing symmetry. It
should be emphasized that the $\text{KFe}_{2}\text{As}_{2}$ is regarded as a
strongly correlated system with strong AF fluctuation evidenced by the
remarkable mass enhancement\cite%
{UjiFSmassEnhance,HardyCorrelationIncoherence2013} and high incoherent
spectral weight\cite{Uchida2014}. Since the main ingredients determining the
pairing symmetry are the FS structure and the electronic interactions
responsible for pairing, it can be concluded that the phase transitions are
essentially driven by the competition between the local spin AF exchanges,
because the FS remains almost the same.

First of all, we identify a plaquette-centered rotation symmetry $C_{4}^{p}$
which is complementary to the $S_{4}$ symmetry, providing a simple and
potentially universal organizing principle for the tangling orbitals of
Fe-based superconductors. We explicitly show that, while $S_{4}$ plays the
role of organizing two groups of orbitals, $C_{4}^{p}$ is responsible for
determining and stabilizing the FS structure of iron-pnictides, including
the heavily hole doped limit where the electron Fermi pockets are absent.
The joint cooperation of the $S_{4}$ and $C_{4}^{p}$ symmetries manifest its
power particularly in settling the controversy of pairing symmetry and
nodeness in KFe$_{2}$As$_{2}$.

To study the pairing symmetry of SC in $\text{KFe}_{2}\text{As}_{2}$, we
take the strong coupling approach by modeling this material with the $%
t-J_{1}-J_{2}$ model subjected to the particle occupancy constraint, in
contrast to the models with the on-site Coulomb repulsion and Hund's coupling%
\cite{Vafek}. Since we are mainly concerned with the pairing symmetry of
superconductivity, we can neglect the quantum fluctuation for the moment and
adopt the slave-boson mean-field approximation. We find that, by decreasing
the ratio of $J_{1}/J_{2}$, the $\text{KFe}_{2}\text{As}_{2}$ samples could
experience phase transitions of pairing symmetry from the $%
d_{x^{2}-y^{2}}\times s_{x^{2}+y^{2}}$-wave SC to $s_{x^{2}y^{2}}$-wave SC
through a narrow intermediate $s+id\times s$ mixed pairing phase. Although
the $d$-wave pairing symmetry is indeed energetically unfavorable compared
with the $s$-wave pairing, the electronic structure imposes an indispensable
$d$-wave form factor on the Cooper pairs glued by the local spin AF coupling
$J_{1}$. This $d$-wave form factor is inherited from the orbital
hybridization in the representation of $C_{4}^{p}$ and becomes the
characteristic nature of multi-orbital iron-pnictide superconductors.
Moreover, together with the $S_{4}$ symmetry, we are in fact bestowed with 2
copies of $d$-wave gap structure, counting 8 nodes in total. By weakly
breaking $C_{4}^{p}$ we are naturally led to the so-called "octet-noded
monster" observed by laser ARPES. We thus propose a symbolic equation "$%
8=4+4 $" that captures the essential physics ruled by $S_{4}$ and $C_{4}^{p}$
symmetries, exhibiting the origin of nodes in a rather simple and
comprehensive way.

Moreover, we find that a different representation of this plaquette-centered
rotation symmetry $\tilde{C}_{4}^{p}$ can stabilize the FS structure of
iron-chalcogenides, including the monolayer FeSe on SrTiO$_{3}$ substrate
where a tiny hole Fermi pocket around $\Gamma $ vanishes\cite%
{XJZhou2012,DLFeng2013}. So both families of Fe-based superconductors share
the plaquette-centered rotation symmetry, and the importance of this
symmetry lies in its capability to understand the band topology and
robustness of FSs, and to reconcile some seemingly contradictory
experimental observations in a rather simple, inevitable, and comprehensive
manner.

\section{Results}

$C_{4}^{p}$ \textbf{symmetry and band topology}

\begin{figure}[t]
\includegraphics[width=7cm]{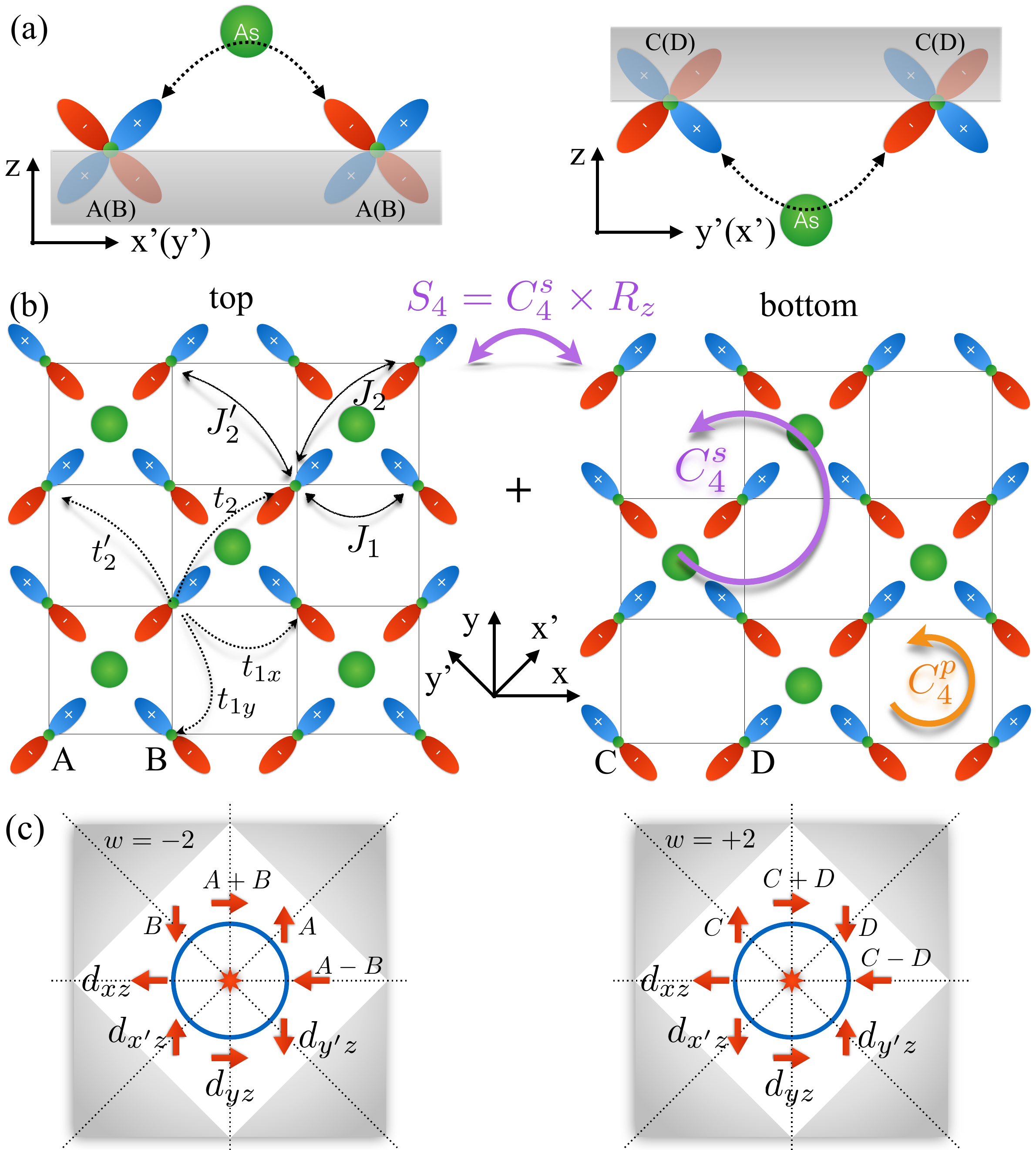}
\caption{Lattice configuration of the iron-pnictides. (a) The transverse
view of the Fe-As plane shows the hopping of $3d_{x^{\prime }(y^{\prime })z}$
orbital via hybridization with $p_{x^{\prime }(y^{\prime })}$ orbital on
arsenic atoms, either above or below the Fe square lattice plane. (b) The $%
3d_{x^{\prime }(y^{\prime })z}$ orbitals on the Fe lattice can be
approximately decoupled into two groups which hop around via the arsenic
atom on top or bottom of the plane respectively. Thus the lattice can be
viewed as "factorized" into the top and bottom layers, which are related by $%
S_{4}$ symmetry. (c) For $d$-wave representation of plaquette-centered
rotation symmetry $C_{4}^{p}$, the $\Gamma $ point is the source of Berry
flux carrying topological winding number $w=\mp 2$ for AB (CD) layer. The
red arrow denotes the polarization of $C_{4}^{p}$ spinor on the Fermi
pocket, where spin up represents A(C) orbitals and spin down represents B(D)
orbitals.}
\label{lattice}
\end{figure}

Due to the weak out-of-plane coupling (along c-axis), the electronic
properties of Fe-based SCs are mainly contained in the FeAs plane, where the
Fe atoms form a square lattice and the As atoms alternate above or below the
Fe plaquette center (Fig.\ref{lattice}a). Because of the checkerboard
pattern of As lattice, the FeAs plane does not respect the site-centered $%
C_{4}^{s}$ symmetry, but is invariant under the site-centered $C_{4}^{s}$
rotation followed by a mirror reflection with respect to the plane\cite%
{S4HuHao2012}: $S_{4}\equiv C_{4}^{s}\times R_{z}$. We further discover that
the plaquette-centered rotation symmetry $C_{4}^{p}$ is preserved.

Among the five $d$-orbitals, the low-energy physics near the FS is mainly
contributed by the $d_{xz},d_{yz},d_{xy}$ orbitals. Instead of a direct
tunneling via the wave function overlap, the hopping between $d_{xz}$- and $%
d_{yz}$-orbitals is primarily contributed by their hybridization with the $p$%
-orbitals on the As atoms, whereas the $d_{xy}$ orbitals do not have this
privilege and are not included in our minimal model. The As atoms on
plaquette centers polarize the $d_{xz}$- and $d_{yz}$-orbitals into $%
d_{x^{\prime }z}$- and $d_{y^{\prime }z}$-orbitals to maximize energy gain
from hopping (Fig.\ref{lattice}a and Fig.\ref{lattice}b). Therefore the
square lattice with $d_{x^{\prime }z}$ and $d_{y^{\prime }z}$ orbitals on
each site is effectively factorized into the top and bottom layers where
each site has one orbital but the unit cell has to be doubled. As shown in
Fig.\ref{lattice}b, the top layer has $d_{x^{\prime }z}$ living on the odd
lattice sites (denoted as A) and $d_{y^{\prime }z}$ on the even lattice
sites (denoted as B). Likewise, the bottom layer has $d_{y^{\prime }z}$
living on the odd lattice sites (denoted as C) and {\ }$d_{x^{\prime }z}$ on
the even lattice sites (denoted as D). For convenience we will denote the
top layer as "AB layer" while the bottom layer as "CD layer". Albeit weakly
coupled by inter-layer tunneling, the two layers are related by the $S_{4}$
symmetry, while the sublattice degrees of freedom inside each layer are
rotated by $C_{4}^{p}$.

To demonstrate the symmetries, we introduce the notion $\tau _{\mu \nu
}\equiv I_{\mu }\otimes I_{\nu }$, $\mu ,\nu =0,1,2,3$, where the first
Pauli matrix $I_{\mu }$ acts on the $S_{4}$ spinor space spanned by those
states living on either layer, and the latter Pauli matrix $I_{\nu }$ acts
on the $C_{4}^{p}$ spinor space composed of states living on each
sublattice. By choosing the simple gauge as shown in Fig.\ref{lattice}b, the
representation of $S_{4}$ and $C_{4}^{p}$ symmetries can be explicitly
expressed as
\begin{eqnarray*}
&&S_{4}:\left(
\begin{array}{c}
\begin{array}{c}
A_{\mathbf{k},\sigma } \\
B_{\mathbf{k},\sigma } \\
C_{\mathbf{k},\sigma } \\
D_{\mathbf{k},\sigma }%
\end{array}%
\end{array}%
\right) \rightarrow \left(
\begin{array}{c}
\begin{array}{c}
C_{\mathbf{k^{\prime }},\sigma } \\
-D_{\mathbf{k^{\prime }},\sigma } \\
-A_{\mathbf{k^{\prime }},\sigma } \\
B_{\mathbf{k^{\prime }},\sigma }%
\end{array}%
\end{array}%
\right) =i\tau _{23}\left(
\begin{array}{c}
\begin{array}{c}
A_{\mathbf{k^{\prime }},\sigma } \\
B_{\mathbf{k^{\prime }},\sigma } \\
C_{\mathbf{k^{\prime }},\sigma } \\
D_{\mathbf{k^{\prime }},\sigma }%
\end{array}%
\end{array}%
\right) , \\
&&C_{4}^{p}:\left(
\begin{array}{c}
\begin{array}{c}
A_{\mathbf{k},\sigma } \\
B_{\mathbf{k},\sigma } \\
C_{\mathbf{k},\sigma } \\
D_{\mathbf{k},\sigma }%
\end{array}%
\end{array}%
\right) \rightarrow \left(
\begin{array}{c}
\begin{array}{c}
B_{\mathbf{k^{\prime }},\sigma } \\
-A_{\mathbf{k^{\prime }},\sigma } \\
-D_{\mathbf{k^{\prime }},\sigma } \\
C_{\mathbf{k^{\prime }},\sigma }%
\end{array}%
\end{array}%
\right) =i\tau _{32}\left(
\begin{array}{c}
\begin{array}{c}
A_{\mathbf{k^{\prime }},\sigma } \\
B_{\mathbf{k^{\prime }},\sigma } \\
C_{\mathbf{k^{\prime }},\sigma } \\
D_{\mathbf{k^{\prime }},\sigma }%
\end{array}%
\end{array}%
\right) ,
\end{eqnarray*}%
where $\mathbf{k^{\prime }}=C_{4}\mathbf{k}$. Since the spin-orbit coupling
is not concerned, the spin degeneracy is always present. The rotation factor
contributed by the spin rotation $(1+i\sigma _{z})/\sqrt{2}$ does not affect
the physics and can be absorbed by a basis transformation. Moreover, we
would like to point out that these two symmetries commute with each other.

Formally, the $C_{4}^{p}$ symmetry operation is indeed diagonal in the $%
S_{4} $ spinor space and mainly rotates the intra-layer sublattices, while
the $S_{4}$ operation is diagonal in each sublattice but primarily rotates
the layers. It is worth noticing that the $S_{4}$ spinor and $C_{4}^{p}$
spinor look like dual to each other. But the inter-sublattice hopping is
much stronger than the inter-layer tunneling, which makes the $S_{4}$
doublet weakly coupled but $C_{4}^{p}$ doublet strongly hybridized. The
power of the $S_{4}$ symmetry lies in that, once the dynamics of one layer
is obtained, it is straightforward to derive the other. Therefore, in the
presence of $S_{4}$ symmetry, we are bestowed with a minimal $C_{4}^{p}$
spinor model living on either layer. In the following we focus on the
properties of the top layer before paying a revisit to the complete $S_{4}$
spinor.

As shown in Fig.\ref{lattice}b, the kinetic part of the model Hamiltonian
mainly involves anisotropic nearest neighbor (NN) hopping $t_{1}$ and the
next nearest neighbor (NNN) hopping $t_{2}$ or $t_{2}^{\prime }$, which can
be recombined and decomposed into $s$-wave and $d$-wave representation: $%
t_{1s}=\left( t_{1x}+t_{1y}\right) /2$, $t_{1d}=\left( t_{1x}-t_{1y}\right)
/2$, $t_{2s}=\left( t_{2}+t_{2}^{\prime }\right) /2$, and $t_{2d}=\left(
t_{2}-t_{2}^{\prime }\right) /2$. In terms of the $C_{4}^{p}$ spinor $\Psi _{%
\mathbf{k},\sigma }\equiv \left( A_{\mathbf{k},\sigma },B_{\mathbf{k},\sigma
}\right) ^{T}$, it is then expressed as
\begin{equation}
H_{t}^{\text{AB}}=\frac{1}{2}\sum_{\mathbf{k},\sigma }\Psi _{\mathbf{k}%
,\sigma }^{\dagger }\left[ \epsilon _{0}(\mathbf{k})+\epsilon _{x}(\mathbf{k}%
)I_{1}+\epsilon _{z}(\mathbf{k})I_{3}\right] \Psi _{\mathbf{k},\sigma },
\end{equation}%
where $\mathbf{k}$ ranges in the unfolded BZ and
\begin{eqnarray*}
\epsilon _{0}(\mathbf{k}) &=&4t_{2s}\cos k_{x}\cos k_{y}-\mu , \\
\epsilon _{z}(\mathbf{k}) &=&-4t_{2d}\sin k_{x}\sin k_{y}, \\
\epsilon _{x}(\mathbf{k}) &=&2t_{1d}\left( \cos k_{x}-\cos k_{y}\right)
+2t_{1s}\left( \cos k_{x}+\cos k_{y}\right) \\
&\equiv &\epsilon _{xd}(\mathbf{k})+\epsilon _{xs}(\mathbf{k}).
\end{eqnarray*}%
Note that $\epsilon _{z}(\mathbf{k})$ denotes the sublattice energy
difference and $\epsilon _{x}(\mathbf{k})$ is the energy gain from the NN
hopping process. A vector can be defined by $\mathbf{h}(\mathbf{k})\equiv
(\epsilon _{x},0,\epsilon _{z})$, which acts as a "magnetic field" in the
momentum space and pinning the $C_{4}^{p}$ spinor (Fig.\ref{lattice}c). It
should be noted that the sublattice degree of freedom in AB layer is locked
with the atomic internal angular momentum i.e. the odd sites carry $%
d_{x^{\prime }z}$ orbitals while the even sites carry $d_{y^{\prime }z}$
orbitals, so that the $C_{4}^{p}$ spinor is actually a composite of the
sublattice degree of freedom and the internal atomic angular momentum. For
instance, when $t_{1s}=0$,$\ t_{1d}<0$,$\ t_{2s}>0$, and$\ t_{2d}>0$, the
spinor along $k_{x}$ with the lowest energy is composed of the equal
superposition of $d_{x^{\prime }z}$-orbitals on odd sites and $d_{y^{\prime
}z}$-orbitals on even sites (Fig.\ref{lattice}c), equivalent to $d_{xz}$%
-orbitals. But along the BZ diagonal, the spinor with the lowest energy
consists of the $d_{x^{\prime }z}$-orbitals on odd sites or $d_{y^{\prime
}z} $-orbitals on even sites.

Before diagonalization, there are several remarks about the symmetries in
this kinetic part. The $d$-wave and $s$-wave NN hopping correspond to two
distinctive symmetry representations of the plaquette-centered rotation
symmetry, respectively. Based on the sketch of a snapshot of the wave
function distribution in Fig.\ref{lattice}b, we naturally expect $%
|t_{1d}|\gg |t_{1s}|\approx 0$. Indeed, adding a nonzero $t_{1s}$ would
break the symmetry $C_{4}^{p}$. So the ideal case of $t_{1s}=0$ and $%
t_{1d}\neq 0$ respects the symmetry $C_{4}^{p}$, which can faithfully
characterize the FS structure of iron-pnictides.
\begin{figure}[t]
\includegraphics[width=7cm]{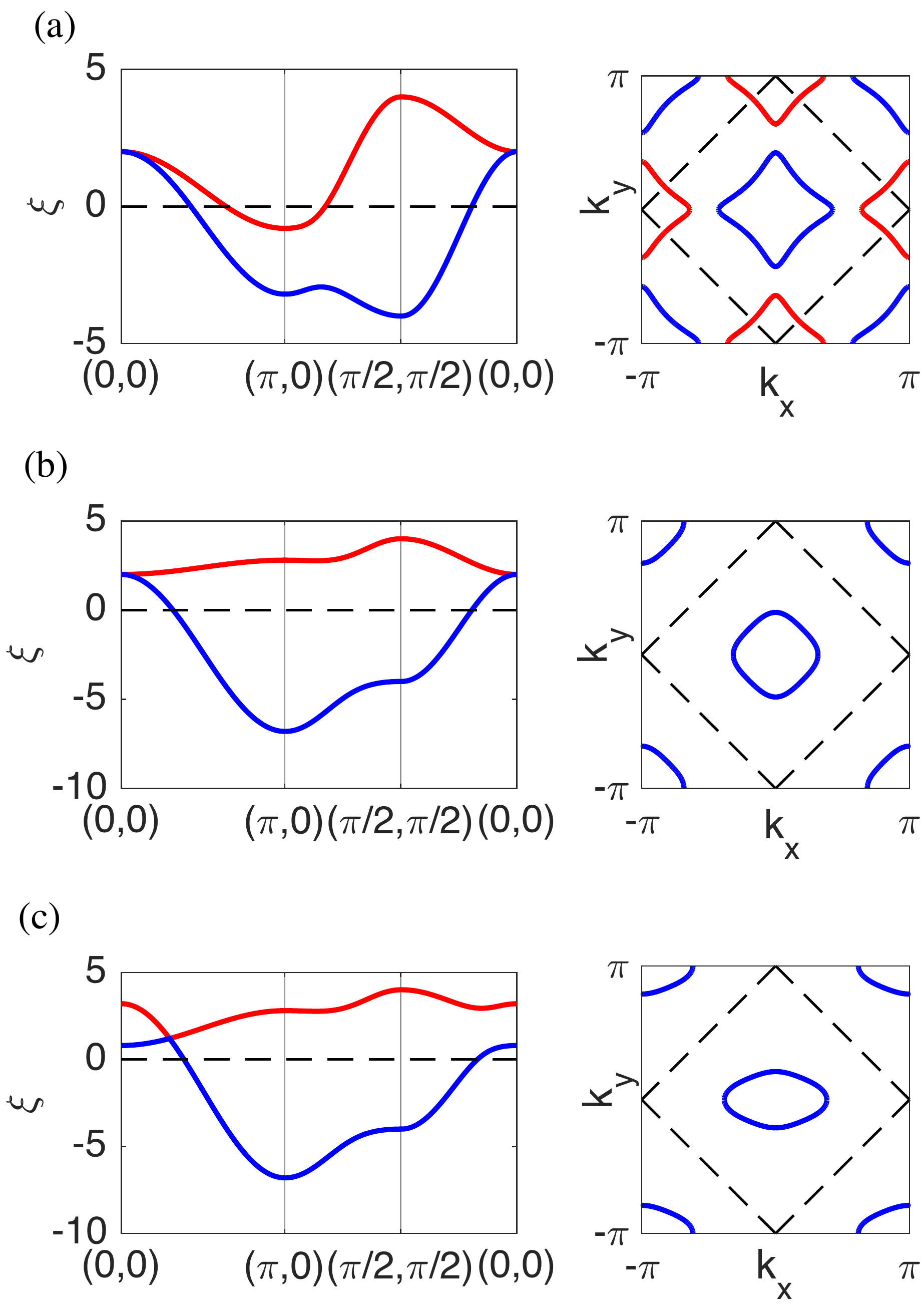}
\caption{Band structures and Fermi surfaces of the minimal model without
considering any electron interactions for various systems of Fe-based
superconductors. Left panel shows the band structure for varying parameters
while right panel shows the corresponding FS in the unfolded BZ. (a) The
parameters are $t_{1d}=-0.3$, $t_{1s}=0$, $t_{2s}=0.5$, in unit of $t_{2d}=1$%
. Hole pocket is located around $\Gamma $ and electron pocket around $X$,
describing the FS structure for iron-pnictides. The $\Gamma $ point is a
double Dirac point with quadratic band touching dispersion. (b) The
parameters are $t_{1d}=-1.2$, $t_{1s}=0$, $t_{2s}=0.5$, in unit of $t_{2d}=1$%
. Electron pocket vanishes, while the hole pocket around $\Gamma $ is
robust, standing for the heavily hole doped iron-pnictides. (c) The
parameters are $t_{1d}=-1.2$, $t_{1s}=-0.3$, $t_{2s}=0.5$, in unit of $%
t_{2d}=1$. Mixing $d$-wave NN hopping with $s$-wave component can split the
double Dirac point into two Dirac points while elongating the hole pocket.
Hole pocket could be further torn apart.}
\label{band}
\end{figure}

In the helicity basis, $H_{t}^{\text{AB}}$ can be directly diagonalized
\begin{equation}
H_{t}^{\text{AB}}=\frac{1}{2}\sum_{\mathbf{k},\sigma }\left[ \xi _{e}(%
\mathbf{k})\alpha _{\mathbf{k},\sigma }^{\dagger }\alpha _{\mathbf{k},\sigma
}+\xi _{h}(\mathbf{k})\beta _{\mathbf{k},\sigma }^{\dagger }\beta _{\mathbf{k%
},\sigma }\right] ,
\end{equation}%
where $\xi _{e(h)}(\mathbf{k})=\epsilon _{0}\pm \sqrt{\epsilon
_{x}^{2}+\epsilon _{z}^{2}}$ represents the electron (hole) band with $\pm $
helicity, respectively:
\begin{eqnarray}
\alpha _{\mathbf{k},\sigma }^{\dagger } &=&\left( \cos \frac{\theta _{%
\mathbf{k}}}{2}\right) A_{\mathbf{k},\sigma }^{\dagger }+\text{sgn}\left(
\epsilon _{x}\right) \left( \sin \frac{\theta _{\mathbf{k}}}{2}\right) B_{%
\mathbf{k},\sigma }^{\dagger },  \notag \\
\beta _{\mathbf{k},\sigma }^{\dagger } &=&\left( \cos \frac{\theta _{\mathbf{%
k}}}{2}\right) B_{\mathbf{k},\sigma }^{\dagger }-\text{sgn}\left( \epsilon
_{x}\right) \left( \sin \frac{\theta _{\mathbf{k}}}{2}\right) A_{\mathbf{k}%
,\sigma }^{\dagger },  \label{hybrid_rel}
\end{eqnarray}%
and the hybridization angle is given by $\theta _{\mathbf{k}}=\tan ^{-1}%
\frac{\left\vert \epsilon _{x}\right\vert }{\epsilon _{z}}\in \left[ 0,\pi %
\right] $. The band dispersion and the corresponding FS structures can be
easily obtained. Fig.\ref{band}a exhibits the coexistence of a hole pocket
around $\Gamma $ and an electron pocket around $X$ point in the absence of
the $s$-wave NN hopping, a characteristic FS structure of most
iron-pnictides. As the $d$-wave NN hopping is growing, the electron band is
gradually pushed upwards, shrinking the electron pocket. It finally leads to
the FS structure of $\text{KFe}_{2}\text{As}_{2}$ (Fig.\ref{band}b), where
the electron pocket completely vanishes. This evolution is also related to
the doping process of Ba$_{1-x}$K$_{x}$Fe$_{2}$As$_{2}$ iron-pnictides\cite%
{Grinenko2017}.

Note that the shape of the hole pocket has the $C_{4}$ rotational symmetry
thanks to the symmetry $C_{4}^{p}$, and the two bands touch each other
quadratically at zone center $\Gamma $ point, which is not accidental but
rather due to the topological nature. Namely, the helical spinor is pinned
by the vector field $\mathbf{h}(\mathbf{k})$ to wind around the $\Gamma $
point twice, yielding a topological number $w=2\text{sgn}\left(
t_{1d}t_{2d}\right) $ (Ref.\cite{RanZhaiVishwanathLee2009}). It is this
topological number that forces the electron and hole bands to touch
quadratically at the vortex core of $\mathbf{h}(\mathbf{k})$, which becomes
the source of the Berry flux experienced by the helical spinor (see Fig.\ref%
{lattice}c). Consequently, the hole pocket surrounding the vortex core is
protected by topology, whereas the electron pocket is less robust. As long
as the time reversal (TR) symmetry is present, the occurrence of $I_{2}$ is
forbidden and this vortex core is robust. However, the TR symmetry cannot
prevent the double Dirac point at $\Gamma $ from splitting apart into two
separated Dirac nodes, which could be driven by a weak $C_{4}^{p}$-breaking
perturbation. This is indeed the case when the $s$-wave hoping comes in
mixing the $d$-wave NN hopping (Fig.\ref{band}c), where the hole pocket
experiences a "nematic force" and is simultaneously elongated. When the $%
C_{4}^{p}$ symmetry breaking term is large enough, the hole pocket around $%
\Gamma $ is going to be torn apart, resulting in two pockets surrounding the
Dirac nodes, which is not seen in the real materials. So the iron-pnictides
with hole pockets around $\Gamma $ robustly observed in experiments must
preserve the $C_{4}^{p}$ symmetry, at least approximately. In other words,
the FS structure of iron-pnictides is stabilized by the $C_{4}^{p}$ symmetry
which helps the TR symmetry protect and confine the double Dirac point to
the hole pocket center. In this sense, the $C_{4}^{p}$ symmetry is one of
the key relevant features of iron-pnictides, which can tolerate only tiny
amount of $s$-wave hopping. In the next section, we will focus on the ideal
case of $d$-wave NN hopping limit by sending $t_{1s}\rightarrow 0$ for $%
\text{KFe}_{2}\text{As}_{2}$. Specifically, we choose the realistic hopping
parameters as $t_{1d}=-1.2$, $t_{2s}=0.5$, and $t_{2d}=1$, whose
corresponding band structure is given by Fig.\ref{band}b. When things are
clear in the ideal case, we will introduce a weak $s$-wave NN hopping to
model the more realistic material later on.

$C_{4}^{p}$ \textbf{protected }$d$\textbf{-wave renormalized pairing SC}

Provided the electronic structure of KFe$_{2}$As$_{2}$, we need to include
the electronic interactions. The strong correlation evidenced by experiments%
\cite{UjiFSmassEnhance,HardyCorrelationIncoherence2013,Uchida2014} requires
a strong coupling approach. Particularly, the optical measurement\cite%
{Uchida2014} had showed the incoherent spectral weight of $\text{KFe}_{2}%
\text{As}_{2}$ as high as 10\% hole-doped $\text{La}_{2-x}\text{Sr}_{x}\text{%
CuO}_{4}$, revealing the strong local AF correlation similar to the
cuprates. Therefore, we would like to model this system with AF
super-exchanges containing the NN and NNN interactions as shown in Fig.\ref%
{lattice}b. Distinct from the $t-J_{1}-J_{2}$ model in the previous form\cite%
{HuBernevig2orbitalExchange}, our model Hamiltonian is subjected to a
constraint that projects out double-occupancy. Thus, the model Hamiltonian
for AB layer is described by $H^{\text{AB}}=\mathcal{P}H_{t}^{\text{AB}}%
\mathcal{P}+H_{J}^{\text{AB}}$ with interaction terms:%
\begin{eqnarray}
H_{J}^{\text{AB}} &=&J_{1}\sum_{\mathbf{r},\mathbf{\delta }}S_{\mathbf{r}%
}^{A}\cdot S_{\mathbf{r}+\mathbf{\delta }}^{B}  \notag \\
&&+J_{2}\sum_{\mathbf{r}}\left( S_{\mathbf{r}}^{A}\cdot S_{\mathbf{r}+\left(
\hat{x}+\hat{y}\right) }^{A}+S_{\mathbf{r}}^{B}\cdot S_{\mathbf{r}+\left(
\hat{x}-\hat{y}\right) }^{B}\right)  \notag \\
&&+J_{2}^{\prime }\sum_{\mathbf{r}}\left( S_{\mathbf{r}}^{A}\cdot S_{\mathbf{%
r}+\left( \hat{x}-\hat{y}\right) }^{A}+S_{\mathbf{r}}^{B}\cdot S_{\mathbf{r}%
+\left( \hat{x}+\hat{y}\right) }^{B}\right) ,
\end{eqnarray}%
where $\mathbf{\delta }=\pm \hat{x},\hat{y}$ is the NN vector. As the
low-energy descendant of the on-site Hubbard interaction, the AF
super-exchange $J$-terms rely on the corresponding hopping integrals. Like
cuprates, the parameters can be chosen as $J_{2}=0.5$ and $J_{2}^{\prime
}=0.2$ as roughly one third of the corresponding hopping integrals and the
dopant concentration is fixed at $0.05$. Note that although $J_{2}^{\prime }$
differs from $J_{2}$, only the combination $J_{2s}\equiv 2J_{2}J_{2}^{\prime
}/(J_{2}+J_{2}^{\prime })$ contributes to pairing. So that the physics is
essentially captured by the competition between $J_{1}$ and $J_{2s}$. The
parameter $J_{1}$ is chosen as the tuning parameter that mimics the inverse
pressure when compared with the pressure experiments\cite{TaftiTaillefer2013}%
.

The projector $\mathcal{P}$ declares the particle occupancy constraints: $%
\sum_{\sigma }A_{\mathbf{r},\sigma }^{\dagger }A_{\mathbf{r},\sigma }\leq 1$
and $\sum_{\sigma }B_{\mathbf{r},\sigma }^{\dagger }B_{\mathbf{r},\sigma
}\leq 1$, which lead to emergent spin-charge separation physics. To tackle
the constraint, we adopt the well-established slave-boson decomposition to
factorize electron into fermionic spinon and bosonic holon:
\begin{equation}
A_{\mathbf{r},\sigma }=h_{\mathbf{r}}^{\dagger }a_{\mathbf{r},\sigma },B_{%
\mathbf{r},\sigma }=h_{\mathbf{r}}^{\dagger }b_{\mathbf{r},\sigma },
\end{equation}%
in which way the constraint becomes equalities: $\sum_{\sigma }a_{\mathbf{r}%
,\sigma }^{\dagger }a_{\mathbf{r},\sigma }+h_{\mathbf{r}}^{\dagger }h_{%
\mathbf{r}}=1\text{ and }\sum_{\sigma }b_{\mathbf{r},\sigma }^{\dagger }b_{%
\mathbf{r},\sigma }+h_{\mathbf{r}}^{\dagger }h_{\mathbf{r}}=1$, and can be
enforced via a Lagrangian multiplier. After the bosonic holons are condensed
$\left\langle h_{\mathbf{r}}^{\dagger }\right\rangle =\left\langle h_{%
\mathbf{r}}\right\rangle =\sqrt{x}$, we introduce the uniform valence bond
and singlet pairing order parameters
\begin{equation*}
\kappa _{ij}=-\frac{J}{4}\left\langle \sum_{\sigma }a_{i,\sigma }^{\dagger
}a_{j,\sigma }\right\rangle ,\Delta _{ij}=\frac{J}{4}\left\langle
a_{i,\uparrow }a_{j,\downarrow }-a_{i,\downarrow }a_{j,\uparrow
}\right\rangle ,
\end{equation*}%
to decouple the local AF interactions. Moreover, the NN and NNN valence bond
and singlet pairing order parameters in real space can be rearranged into
the $s$-wave and $d$-wave representations: $\kappa _{1}$ is automatically of
$d$-wave symmetry required by $C_{4}^{p}$ symmetry and other valence bond
orders are%
\begin{eqnarray}
\kappa _{2s} &=&(\kappa _{2}+\kappa _{2}^{\prime })/2,\ \kappa _{2d}=(\kappa
_{2}-\kappa _{2}^{\prime })/2,\   \notag \\
\Delta _{1s} &=&(\Delta _{1x}+\Delta _{1y})/2,\ \Delta _{1d}=(\Delta
_{1x}-\Delta _{1y})/2,\   \notag \\
\Delta _{2s} &=&(\Delta _{2}+\Delta _{2}^{\prime })/2,\Delta _{2d}=(\Delta
_{2}-\Delta _{2}^{\prime })/2.
\end{eqnarray}

Finally, a mean-field Hamiltonian can be obtained
\begin{equation}
H_{\text{MF}}^{\text{ab}}=H_{t}^{\text{ab}}+H_{\Delta }^{\text{ab}},
\end{equation}%
which describes the superconducting quasiparticles on the AB layer. The
kinetic part $H_{t}^{\text{ab}}$ has the same band structure as we discussed
in the last section. $H_{t}^{\text{AB}}\rightarrow H_{t}^{\text{ab}}$
requires replacing the electrons with the corresponding quasiparticles $\Psi
_{\mathbf{k},\sigma }=\left( A_{\mathbf{k},\sigma },B_{\mathbf{k},\sigma
}\right) ^{T}\rightarrow \psi _{\mathbf{k},\sigma }\equiv \left( a_{\mathbf{k%
},\sigma },b_{\mathbf{k},\sigma }\right) ^{T}$, while renormalizing the
hopping integrals and the chemical potential as
\begin{eqnarray}
t_{1d} &\rightarrow &\tilde{t}_{1d}=\left( t_{1d}x+\kappa _{1}\right) ,\text{
}t_{1s}\rightarrow \tilde{t}_{1s}=t_{1s}x,  \notag \\
t_{2s} &\rightarrow &\tilde{t}_{2s}=\left( t_{2s}x+\kappa _{2s}\right) ,
\notag \\
t_{2d} &\rightarrow &\tilde{t}_{2d}=\left( t_{2d}x+\kappa _{2d}\right) ,
\notag \\
\mu _{0} &\rightarrow &\mu =\mu _{0}-\lambda -\left( J_{2}+J_{3}\right) /4.
\end{eqnarray}%
The hybridization relation Eq. (\ref{hybrid_rel}) also holds by replacing
A(B) with a(b), and the mean-field pairing terms can be compactly expressed
in terms of the $C_{4}^{p}$ spinor:
\begin{equation}
H_{\Delta }^{\text{ab}}=\frac{1}{2}\sum_{\mathbf{k}}\left[ \psi _{\mathbf{k}%
,\uparrow }^{\dagger }\left( \Delta _{0}+i\Delta _{x}I_{x}+\Delta
_{z}I_{z}\right) (\psi _{\mathbf{k},\downarrow }^{\dagger })^{T}+h.c.\right]
\end{equation}%
with
\begin{eqnarray*}
\Delta _{0}(\mathbf{k}) &=&4\Delta _{2s}\cos k_{x}\cos k_{y},\Delta _{z}(%
\mathbf{k})=4\Delta _{2d}\sin k_{x}\sin k_{y}, \\
\Delta _{x}(\mathbf{k}) &=&2\Delta _{1s}\left( \cos k_{x}+\cos k_{y}\right)
+i2\Delta _{1d}\left( \cos k_{x}-\cos k_{y}\right) .
\end{eqnarray*}%
Provided with all the form factors available from the interactions, we can
eliminate some of them which are apparently unfavorable energetically.

In this minimal model for KFe$_{2}$As$_{2}$, within the AB layer there is
only a small hole pocket around $\Gamma $ point. The Cooper pair formed on
the FS is more likely to be scattered onto the same FS with small momentum
transfer. And the pairing interactions $J_{1}(q)\propto -2J_{1}\left( \cos
q_{x}+\cos q_{y}\right) $ and $J_{2s}(q)\propto -4J_{2s}\cos q_{x}\cos q_{y}$
are attractive when $q\approx 0$, while $J_{2d}(q)\propto 4J_{2d}\sin
q_{x}\sin q_{y}$ tends to vanish. Therefore, the pairing components $\Delta
_{1d}$ and $\Delta _{2d}$ are energetically unfavorable compared to $\Delta
_{1s}$ and $\Delta _{2s}$, which endow the FS with largest possible energy
gain. In fact, we did some self-consistent calculation numerically to verify
the results $\Delta _{1d}=\Delta _{2d}=0$. Thus we are left with two $s$%
-wave pairing components $\Delta _{1s}$ and $\Delta _{2s}$ to be determined
self-consistently by minimizing the ground state energy.

As there is only single hole pocket FS in the low energy excitations, we can
project the effective Hamiltonian onto the hole band and especially focus on
the vicinity of FS. Turned to the band basis $\Gamma _{\mathbf{k},\sigma
}\equiv \left( \alpha _{\mathbf{k},\sigma },\beta _{\mathbf{k},\sigma
}\right) ^{T}$, the mean-field Hamiltonian can be straightforwardly obtained
\begin{eqnarray}
&&H_{\text{eff}}^{\text{ab}}=\frac{1}{2}\sum_{\mathbf{k},\sigma }\xi _{h}(%
\mathbf{k})\beta _{\mathbf{k},\sigma }^{\dagger }\beta _{\mathbf{k},\sigma }
\\
&&+\frac{1}{2}\sum_{\mathbf{k}}\left( \Delta _{0}(\mathbf{k})-\frac{%
i\epsilon _{xd}(\mathbf{k})\Delta _{x}(\mathbf{k})}{\sqrt{\epsilon
_{xd}^{2}+\epsilon _{z}^{2}}}\right) \beta _{\mathbf{k},\uparrow }^{\dagger
}\beta _{-\mathbf{k},\downarrow }^{\dagger }+h.c..  \notag
\end{eqnarray}%
Here we can see that the $s$-wave pairing component $\Delta _{x}$ arising
from the NN spin exchange interaction is renormalized by a $d$-wave form
factor $\epsilon _{xd}\propto \left( \cos k_{x}-\cos k_{y}\right) $, which
is inherited from the NN hopping integral in the $C_{4}^{p}$ symmetry
representation. Actually it should be no surprise, since we can physically
understand this renormalization factor in the following way. The NN pairing
interaction glues the inter-sublattice particles, which coexist in the hole
band with probability proportional to the hybridization energy gain. As a
result, the effective intra-hole-band pairing condensate is supposed to be
renormalized by $\epsilon _{xd}$. As shown in Fig.\ref{lattice}c, along the
zero lines of hybridization energy $\epsilon _{xd}$, there is no coexistence
of the sublattice degrees of freedom so that there are no inter-orbital
Cooper pairs, regardless of their internal pairing symmetry. Such a
mechanism is parallel to the effective intra-band $p$-wave pairing induced
on the helical electrons in proximity to the $s$-wave superconductors\cite%
{FuKane2008}.

Then the Bogoliubov quasi-particle spectrum can be derived
\begin{equation}
E(\mathbf{k})=\pm \sqrt{\xi _{h}^{2}\left( \mathbf{k}\right) +\Delta
_{0}^{2}\left( \mathbf{k}\right) +\tilde{\Delta}_{x}^{2}(\mathbf{k})},
\end{equation}%
where $\tilde{\Delta}_{x}(\mathbf{k})\equiv \frac{\epsilon _{xd}(\mathbf{k})%
}{\sqrt{\epsilon _{xd}^{2}+\epsilon _{z}^{2}}}\Delta _{x}(\mathbf{k})$ and $%
E(\mathbf{k})$ exhibits the nodal excitations when $\Delta _{0}\left(
\mathbf{k}\right) =0$. Note that the inter-band pairing between hole band
and high energy electron band only contributes second order perturbation
corrections to the gap structure, but is unable to alter its symmetry. The $%
d $-wave gap nodes carrying one unit of vorticity in the Nambu space are
topologically protected\cite{FaDunghai,eDopedZhuZhang}. Within the AB layer,
there are only two ways to destroy these nodes. The first one is to generate
a mass upon the massless Bogoliubov quasi-particles\cite{ZhuWangZhang} which
is forbidden by the TR symmetry in pairing sector. The mass term is an
additional pairing component with phase difference that cannot be gauged
away. The other way is to move the nodes with opposite vorticity to
annihilate each other, which is prevented by the $C_{4}^{p}$ symmetry that
confines the nodes to the unfolded BZ diagonal.

\begin{figure}[t]
\includegraphics[width=7.8cm]{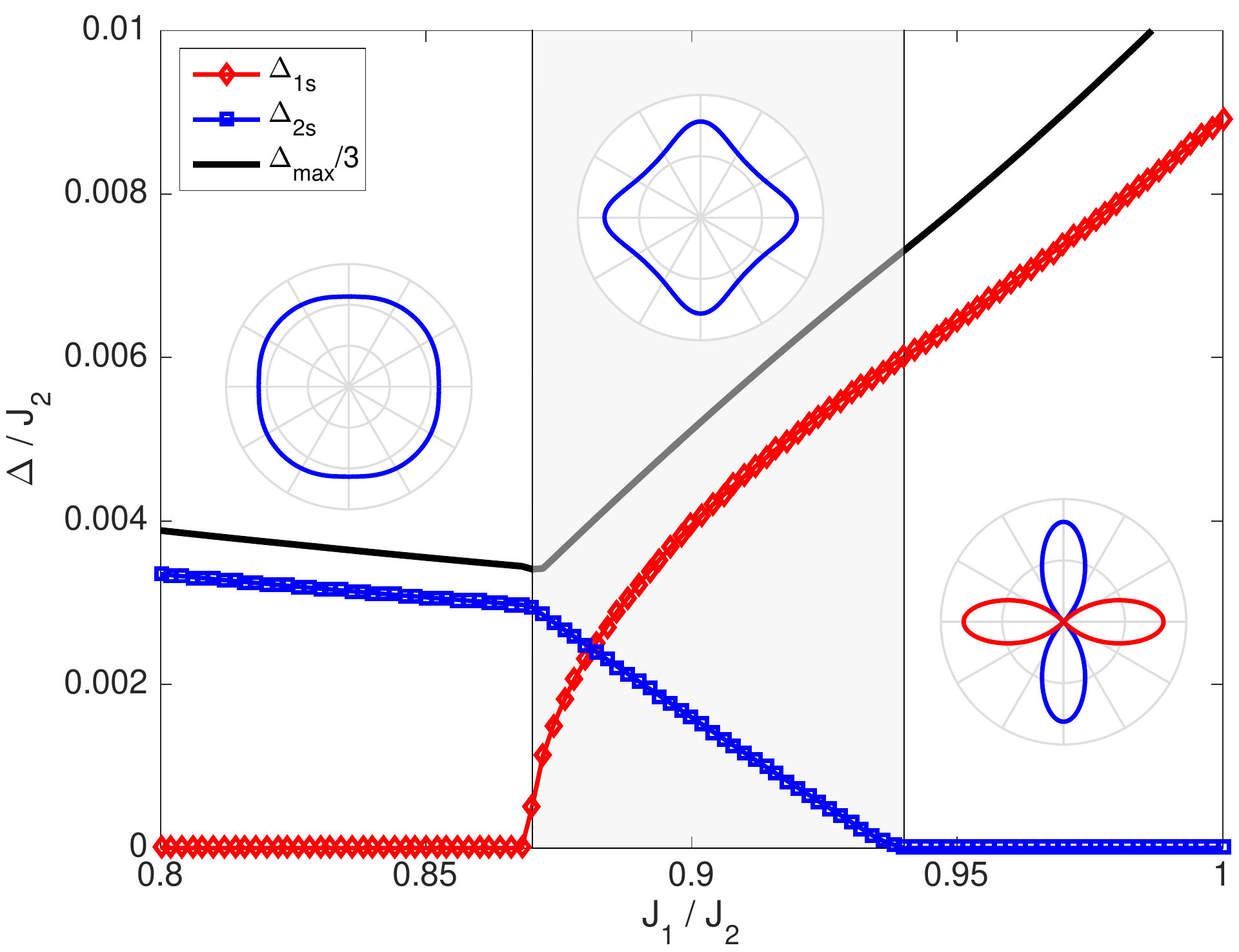}
\caption{Phase diagram of the superconducting phases for KFe$_{2}$As$_{2}$
under pressure (decreasing $J_{1}$ mimics the trend of increasing pressure).
The parameters are $x=0.05$, $t_{1d}=-1.2$, $t_{2s}=0.5$, $J_{2}=0.5$, $%
J_{2}^{\prime }=0.2$ in unit of $t_{2d}=1$. Red line is for the NN Cooper
pair $|\Delta _{1s}|$ that obtains $s_{x^{2}y^{2}}$ form factor in momentum
space, while blue line is for the NNN Cooper pair condensate $|\Delta _{2s}|$
who is endowed with $d_{x^{2}-y^{2}}\times s_{x^{2}+y^{2}}$-wave form
factor. The $d_{x^{2}-y^{2}}\times s_{x^{2}+y^{2}}$-wave SC comes to replace
the $s_{x^{2}y^{2}}$-wave SC when $J_{1}$ overwhelms $J_{2s}$. The grey
color region marks the intermediate phase region with $%
s_{x^{2}y^{2}}+id_{x^{2}-y^{2}}\times s_{x^{2}+y^{2}}$ mixed pairing SC.
Black solid line shows the maximal gap along the FS, which is roughly
proportional to the critical temperature when superfluid density does not
vary drastically. Insets show the pairing gap structure on the FS,
characteristic of the three pairing phases, respectively. }
\label{phase}
\end{figure}

Now we have two competing pairing components: the NNN Cooper pair condensate
$\Delta _{0}(\mathbf{k})$ with the $s_{x^{2}y^{2}}$-wave form factor and the
NN Cooper pair condensate $\tilde{\Delta}_{x}(\mathbf{k})$ carrying the $%
d_{x^{2}-y^{2}}\times s_{x^{2}+y^{2}}$-wave form factor. They can be varied
by the NN interaction $J_{1}$ for a fixed NNN interaction $J_{2s}\simeq 0.3$%
. The detailed self-consistency calculation explicitly shows phase
transitions of pairing symmetry displayed in Fig.\ref{phase}. When $J_{1}\ll
J_{2s}$, the NNN interaction overwhelms the NN interaction, leading to the $%
s_{x^{2}y^{2}}$-wave pairing, whereas $J_{1}\gg J_{2s}$ results in the $%
d_{x^{2}-y^{2}}\times s_{x^{2}+y^{2}}$-wave pairing. In between, the SC with
a mixed pairing $s+id\times s$ is energetically more favorable, which
spontaneously breaks the TR symmetry. To compare with the pressure
experiments\cite{TaftiTaillefer2013}, we notice that, upon increasing
pressure, both $t_{2}$ and $J_{2}$ are expected to grow faster than $t_{1}$
and $J_{1}$. Since the dopant concentration is fixed and the small hole
pocket FS does not change qualitatively, we expect that decreasing the ratio
of $J_{1}/J_{2s}$ is adequate to capture the essential physics during the
period of increasing pressure. When the superfluid density does not vary
drastically, the superconducting critical temperature would be roughly
proportional to the maximum gap on the FS, and it turns out the tendency of
maximum gap on the FS along $J_{1}$ shown by the black solid line in Fig.\ref%
{phase} concurs qualitatively well with the $T_{c}$ trend under decreasing
pressure in experiments\cite{TaftiTaillefer2013}. Moreover, the phase
transition from the $d\times s$-wave nodal SC to the $s+id\times s$ nodeless
SC belongs to the TR breaking mass generation scenario of destroying the
pairing gap nodes.

For the CD layer, we can simply apply the $S_{4}$ symmetry to obtain its
low-energy effective Hamiltonian:
\begin{eqnarray}
&&H_{\text{eff}}^{\text{cd}}=\frac{1}{2}\sum_{\mathbf{k},\sigma }\xi _{h}(%
\mathbf{k})\gamma _{\mathbf{k},\sigma }^{\dagger }\gamma _{\mathbf{k},\sigma
} \\
+ &&\frac{\eta }{2}\sum_{\mathbf{k}}\left( \Delta _{0}(\mathbf{k})+\frac{%
\epsilon _{xd}(\mathbf{k})}{\sqrt{\epsilon _{xd}^{2}+\epsilon _{z}^{2}}}%
i\Delta _{x}(\mathbf{k})\right) \gamma _{\mathbf{k},\uparrow }^{\dagger
}\gamma _{-\mathbf{k},\downarrow }^{\dagger }+h.c.  \notag
\end{eqnarray}%
where $\gamma _{\mathbf{k},\sigma }^{\dagger }=\left( \cos \frac{\theta _{%
\mathbf{k}}}{2}\right) c_{\mathbf{k},\sigma }^{\dagger }-\text{sgn}\left(
\epsilon _{xd}\right) \left( \sin \frac{\theta _{\mathbf{k}}}{2}\right) d_{%
\mathbf{k},\sigma }^{\dagger }$. $\eta $ embodies the global phase
difference between the pairing condensates of the two layers, and is
restricted to be either $\pm 1$ or $\pm i$ as the one-dimensional
representations of the $S_{4}$ point group\cite{S4HuHao2012}. The relative
phase difference can be pinned down by inter-layer couplings for the benefit
of energetics.

\textbf{Octet nodal SC from distorting }$d$\textbf{-wave pairing}

\begin{figure}[t]
\includegraphics[width=7.8cm]{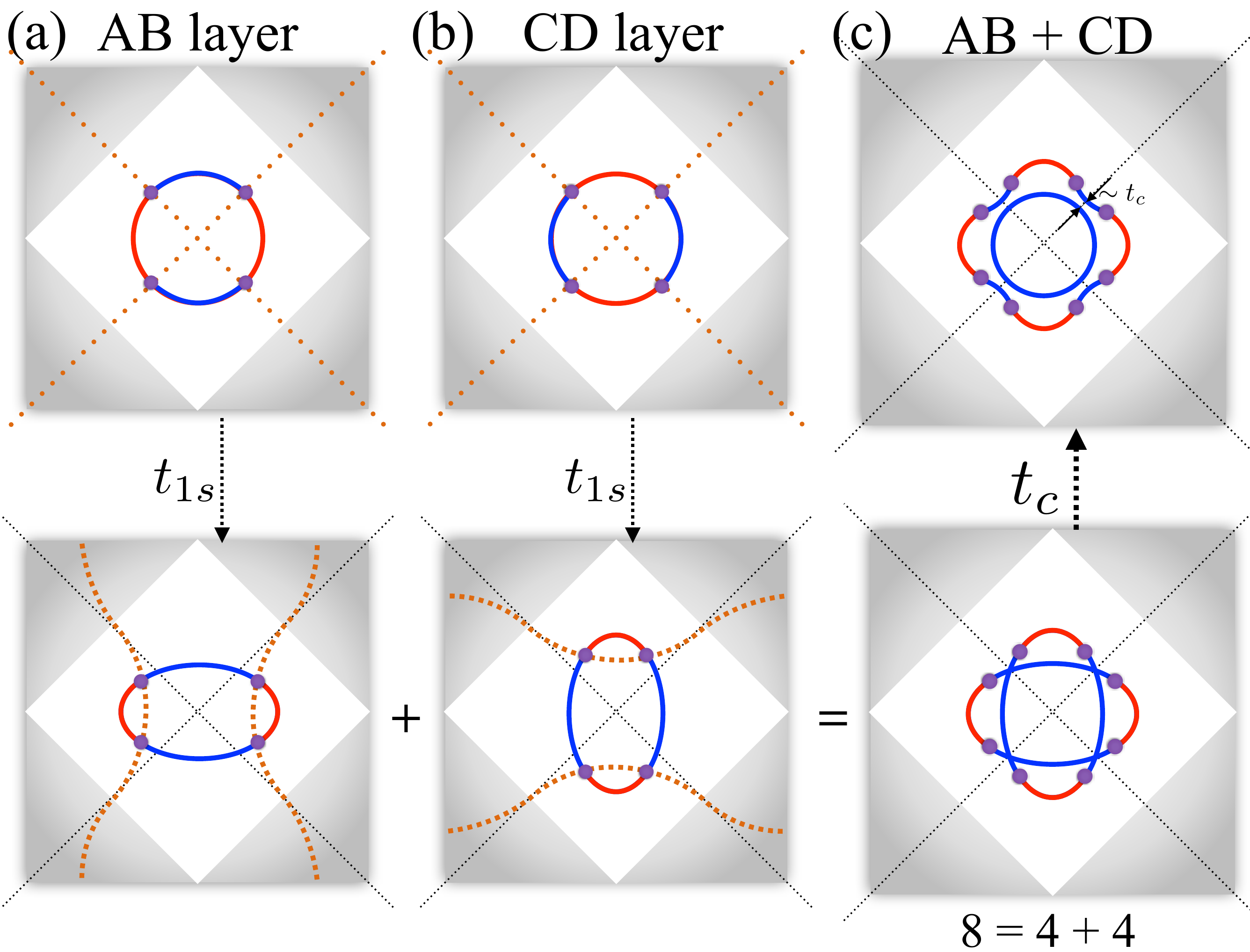}
\caption{Distorted $d$-wave nodes on the Fermi surfaces. The white diamond
emphasizes the folded BZ; the orange dashed lines denote the nodal lines of
the effective $d_{x^{2}-y^{2}}$-wave pairing condensate on the FS distorted
by $s_{x^{2}+y^{2}}$-wave factor (the effect of weak $s$-wave is moderately
exaggerated for illustration); The FS is separated by pairing nodal lines
into the segments with positive (negative) pairing condensate marked by red
(blue) color, respectively. (a) In the AB layer, the mixing weak $s$-wave NN
hopping acts like a "nematic force" that elongates the FS and distorts the $d
$-wave nodal lines, leading to shifted quartet nodes. (b) As the CD layer is
related to AB layer by the $S_{4}$ symmetry, the distortion occurs in the
perpendicular direction. (c) Near unfolded BZ diagonal the quasi-particles
from the AB and CD layers are strongly hybridized with each other by any
infinitesimal inter-layer tunneling. The intersecting elliptic hole pockets
are therefore reconstructed into inner and outer pockets. The inner one has
nodeless pairing gap whereas the outer one shows octet nodal gap structure.
The origin of nodes can be attributed to a rather simple symbolic equation
"8=4+4".}
\label{nodes}
\end{figure}

We are now in a good position to focus on the nodal phase at ambient
pressure, which appears like a sphinx in a variety of experiments. Based on
what we obtained, we have two $S_{4}$-related $C_{4}^{p}$ symmetric hole
pockets which are absolutely degenerate. They can form $d$-wave pairing
condensates with degenerate quartet nodes residing on the unfolded BZ
diagonal (Fig.\ref{nodes}a and \ref{nodes}b). Such degeneracy is highly
unstable against any arbitrary perturbation, so we are obliged to return to
the more realistic materials by involving the weak $C_{4}^{p}$-breaking
hopping $t_{1s}$ and inter-layer tunneling $t_{c}$. As for the inter-layer
tunneling, the $s$-wave NN tunneling would be the dominant term, from the
estimate of orbitals overlap and symmetry analysis (see Fig.\ref{lattice}b).
Note that the on-site tunneling is suppressed because of the orbital
orthogonality. Surprisingly, the $S_{4}$ symmetry survives this tunneling
process. When $t_{c}\ll t_{1s}\ll t_{1d}$, the effective Hamiltonians for
the AB and CD layers are modified by $t_{1s}$ in their respective ways:%
\begin{eqnarray}
H_{\text{eff}}^{\text{ab}} &\rightarrow &\frac{1}{2}\sum_{\mathbf{k},\sigma
}\xi _{h}^{+}(\mathbf{k})\beta _{\mathbf{k},\sigma }^{\dagger }\beta _{%
\mathbf{k},\sigma } \\
&&+\frac{1}{2}\sum_{\mathbf{k}}\frac{\epsilon _{xd}(\mathbf{k})+\epsilon
_{xs}(\mathbf{k})}{\sqrt{\left( \epsilon _{xd}+\epsilon _{xs}\right)
^{2}+\epsilon _{z}^{2}}}\Delta _{x}(\mathbf{k})\beta _{\mathbf{k},\uparrow
}^{\dagger }\beta _{-\mathbf{k},\downarrow }^{\dagger }+h.c.  \notag \\
H_{\text{eff}}^{\text{cd}} &\rightarrow &\frac{1}{2}\sum_{\mathbf{k},\sigma
}\xi _{h}^{-}(\mathbf{k})\gamma _{\mathbf{k},\sigma }^{\dagger }\gamma _{%
\mathbf{k},\sigma } \\
&&+\frac{\eta }{2}\sum_{\mathbf{k}}\frac{-\epsilon _{xd}(\mathbf{k}%
)+\epsilon _{xs}(\mathbf{k})}{\sqrt{\left( \epsilon _{xd}-\epsilon
_{xs}\right) {}^{2}+\epsilon _{z}^{2}}}\Delta _{x}(\mathbf{k})\gamma _{%
\mathbf{k},\uparrow }^{\dagger }\gamma _{-\mathbf{k},\downarrow }^{\dagger
}+h.c.  \notag
\end{eqnarray}%
where the global phase of $\Delta _{1s}$ has been gauged to absorb the phase
factor $-i$. Meanwhile, the normal state dispersion and the hybridization
angle for AB/CD layer are given by
\begin{equation}
\xi _{h}^{\pm }(\mathbf{k})=\epsilon _{0}(\mathbf{k})-\sqrt{\left( \epsilon
_{xd}\pm \epsilon _{xs}\right) ^{2}+\epsilon _{z}^{2}},
\end{equation}%
and $\theta ^{\pm }=\tan ^{-1}\frac{\left\vert \epsilon _{xd}\pm \epsilon
_{xs}\right\vert }{\epsilon _{z}}$, so that the quasiparticles, $\beta _{%
\mathbf{k},\sigma }^{\dagger }$ and $\gamma _{\mathbf{k},\sigma }^{\dagger }$
are adapted (see Supplementary Material for detail). As a result, the $d$%
-wave nodal lines that used to lie across the unfolded BZ diagonal are now
twisted and dragged away from the $\Gamma $ point by a "nematic force",
resulting in a simultaneous movement of nodes along the elongating direction
of the hole Fermi pockets (Fig.\ref{nodes}a and \ref{nodes}b). Since the
nodes with opposite vorticity need to travel a finite path along the FS
before annihilation, they can survive a finite weak $C_{4}^{p}$-breaking
term depending on the size of the FS.

The hole pockets from the AB and CD layers are required by $S_{4}$ symmetry
to intersect at the unfolded BZ diagonal (Fig.\ref{nodes}c). Degeneracy at
this point is unstable against any arbitrarily weak perturbation of
inter-layer tunneling:
\begin{eqnarray}
H_{c} &=&\frac{1}{2}\sum_{\mathbf{k},\sigma }\left[ \epsilon _{c}(\mathbf{k}%
)\left( a_{\mathbf{k},\sigma }^{\dagger }d_{\mathbf{k},\sigma }+b_{\mathbf{k}%
,\sigma }^{\dagger }c_{\mathbf{k},\sigma }\right) +h.c.\right]  \notag \\
&\rightarrow &\frac{1}{2}\sum_{\mathbf{k},\sigma }\left[ \tilde{\epsilon}%
_{c}(\mathbf{k})\beta _{\mathbf{k},\sigma }^{\dagger }\gamma _{\mathbf{k}%
,\sigma }+h.c.\right] .
\end{eqnarray}%
So the degenerate Fermi-points at the unfolded BZ diagonal are split by $%
\tilde{\epsilon}_{c}(\mathbf{k})$ into two Fermi-points, whose corresponding
quasi-particles are given by the bonding and anti-bonding states of the two
layers. The split Fermi points smoothly join the Fermi sheets far away from
the unfolded BZ diagonal, where $\tilde{\epsilon}_{c}(\mathbf{k})$ amounts
only up to second order perturbation corrections (Fig.\ref{nodes}c). Next
let's consider how the pairing matrix changes with the inter-layer
tunneling. It is expected that $\eta =1$ to avoid the destructive
interference along the BZ diagonal where two layers strongly hybridize.
Therefore, the pairing condensates on the unfolded BZ diagonal were
identical for hole pockets from both layers, guaranteed by the $S_{4}$
symmetry (Fig.\ref{nodes}c). The bonding of the two layers does not affect
the pairing matrix along unfolded BZ diagonal, which maintains diagonal in
the form of identity (see Supplementary Material for detail). As a result,
the two reconstructed bands yield the outer and inner hole pockets and are
decoupled in pairing sector to the leading order. The outer pocket inherits
the total octet nodes (Fig.\ref{nodes}c), which still carry vorticity in the
Nambu space and are protected by topology. On the inner pocket the pairing
gap is nodeless. It should be made clear that our outer pocket corresponds
to the "middle pocket" termed in experimental report\cite{OctetOkazaki2012},
because the $d_{xy}$ orbital has been neglected in our minimal model.

In this picture, we have seen how the octet nodes come out naturally as
observed in the laser ARPES\cite{OctetOkazaki2012}, and we can understand
why every two of the octet nodes are located so close to the unfolded BZ
diagonal in experiment, because they are essentially born of the $d$-wave
representation of $C_{4}^{p}$ albeit distorted by the weak $C_{4}^{p}$%
-breaking term. Therefore, the so-called "octet-noded monster" is neither
accidental nor some crazy Cooper pair of angular momentum as high as $g$%
-wave, but the combination of two distorted $d$-wave related by the $S_{4}$
symmetry. In short, the origin of nodes can be attributed to a symbolic
equation "$8=4+4$". The nodes share the same fate with the FS structure that
protests against strong breaking of $C_{4}^{p}$ symmetry. Therefore, we
settle the disagreement between $d$-wave gap structure and the observation
of octet nodes on one of the hole pockets. All in all, the key feature lies
in the multi-orbital character.

\section{Discussion}

\begin{figure}[h]
\includegraphics[width=7.8cm]{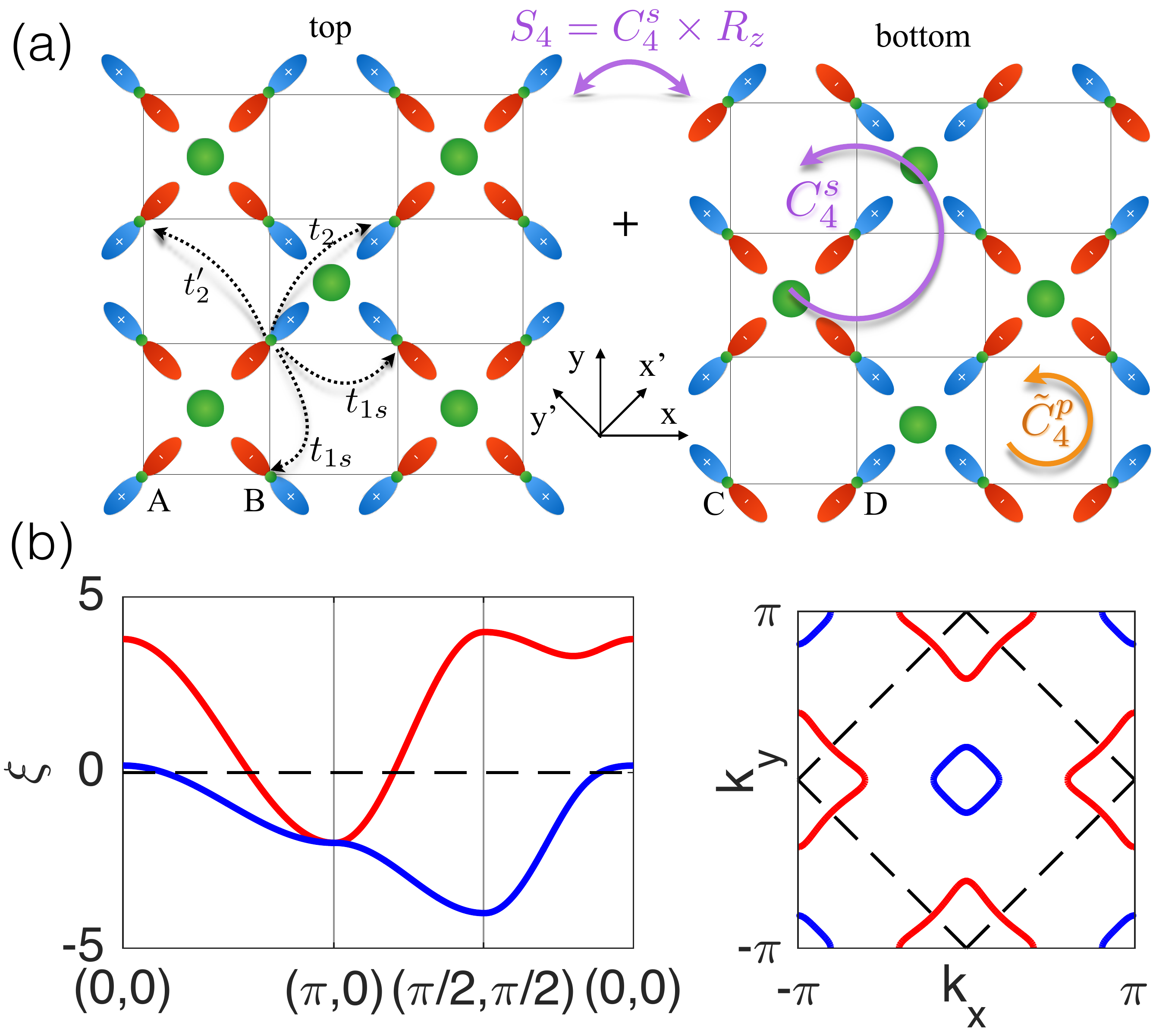}
\caption{(a) Gauge choice characteristic of FeSe, followed by a
particle-hole transform. (b) Band dispersion typical of iron-chalcogenides,
where the electron pocket around $(\protect\pi ,0)$ is robust while hole
pocket around (0,0) is dispensable. The parameters for the demonstration are
chosen to be $t_{1d}=0$, $t_{1s}=0.45$, $t_{2s}=0.5$ in unit of $t_{2d}=1$. }
\label{FeSe}
\end{figure}

Inspired by this organizing principle governed by $C_{4}^{p}$ and $S_{4}$
symmetries, we can gain some insight into the electronic structure and
pairing symmetry of iron-selenide (FeSe). While the $d$-wave representation $%
C_{4}^{p}$ of plaquette-centered rotation symmetry stabilizes the FS
structure of iron-pnictides, the $s$-wave representation $\tilde{C}_{4}^{p}$
captures the key feature of FSs in FeSe as shown in Fig.\ref{FeSe}a, where
the NN $s$-wave hopping dominates. In Fig.\ref{FeSe}b, it can be shown that,
when the $d$-wave NN hopping is completely replaced by the $s$-wave hoping
followed by a particle-hole transform, we can obtain the FS with a robust
electron pocket around $X$ point without hole pocket around $\Gamma $ point,
which is characteristic of FeSe\cite{ZXShenFeSe2016}. In this case, the
electron pocket and its central double Dirac point are protected by the TR
symmetry together with the $s$-wave representation of the plaquette-centered
rotation symmetry $\tilde{C}_{4}^{p}\equiv \tau _{31}$. Thus, the band
structure of FeSe is related to that of iron-pnictides by a gauge transform
combined with a particle-hole transform.

When the pairing symmetry is considered, we can just focus on the monolayer
FeSe with only electron Fermi pockets around $X$ point\cite{ZXShenFeSe2016}.
Parallel to the above discussion for $\text{KFe}_{2}\text{As}_{2}$, there
are two possible pairing symmetries for the monolayer FeSe: one is the $%
s_{x^{2}y^{2}}$-wave Cooper pair glued by the dominant NNN AF coupling $%
J_{2} $ and the other one is the effective $s_{x^{2}+y^{2}}\times
d_{x^{2}-y^{2}}$-wave Cooper pair on the electron pockets around the $X$
point for a relatively large NN AF coupling $J_{1}$. Although the
inter-orbital Cooper pairs glued by $J_{1}$ has $d_{x^{2}-y^{2}}$-wave
symmetry, whose nodal line avoids the FS to maximize the energy gain, the
projection of the inter-orbital Cooper pairs onto the FS yields an
additional $s_{x^{2}+y^{2}}$ form factor inherited from the electronic
structure. However, when $J_{2}$ overwhelms $J_{1}$, the pairing symmetry is
the $s_{x^{2}y^{2}}$-wave. Without more interactions, the two layers
tolerating weak inter-layer tunneling tend to lock their respective phases
of Cooper pairs to be identical, otherwise there would be destructive
interference. This supports the plain s-wave pairing symmetry as observed by
the STM measurement\cite{DLFengPlainS}. But when $J_{1}$ dominates over $%
J_{2}$, the effective $s_{x^{2}+y^{2}}\times d_{x^{2}-y^{2}}$-wave pairing
condensate would show a gap minimum near the folded BZ boundary and a gap
maximum along the unfolded BZ boundary, which seems to concur with the
superconducting gap anisotropy measured by ARPES\cite{ZXShenFeSe2016}.

\begin{acknowledgments}
\textbf{Acknowledgment}\newline
{GMZ acknowledges the support of National Key Research and Development
Program of China (2016YFA0300300).}
\end{acknowledgments}

\textbf{Author contributions}\newline
{GMZ initiated and supervised this project, GYZ conducted the derivation and
calculation, and GYZ and GMZ wrote the paper.}

\end{document}